# Magnetic guidance of charged particles

Dirk Dubbers

*Physikalisches Institut der Universität, Im Neuenheimer Feld 226, 69120 Heidelberg, Germany*
*E-mail address:* dubbers@physi.uni-heidelberg.de

15 July 2015

ABSTRACT

  Many experiments and devices in physics use static magnetic fields to guide charged particles from a source onto a detector, and we ask the innocent question: What is the distribution of particle intensity over the detector surface? One should think that the solution to this seemingly simple problem is well known. We show that, even for uniform guide fields, this is not the case, and we present analytical point spread functions (PSF) for magnetic transport that deviate strongly from previous results. The "magnetic" PSF shows unexpected singularities, which were recently also observed experimentally, and which make detector response very sensitive to minute changes of position, field amplitude, or particle energy. In the field of low-energy particle physics, these singularities may become a source of error in modern high precision experiments, or may be used for instrument tests.

# 1. Introduction

The motion of charged particles in magnetic fields is a highly developed subject, treated in numerous papers and books. The most frequently investigated case is magnetic focussing, as used in electron microscopes, oscilloscopes, electron spectrometers, particle accelerators, or in mass spectrometers with magnetic sector fields. Electron optics was developed early in the past century [1], mainly for small angular ranges of particle emission $\Delta\theta \ll 1$, and for trajectories that describe less than one full orbit of gyration ($n < 1$), see also [2] and the books quoted therein. The case of a magnetic $\beta$-ray spectrometer based on one full orbit ($n = 1$) was treated in [3,4], while a survey on magnetic electron and ion spectrometers is given in [5].

In more recent times, magnetic fields are increasingly being used to simply guide charged particles, like electrons, muons, ions, or other, efficiently from a source to a detector. Such setups are found in magnetic photoelectron imaging [6,7,8], invented in the early 1980ies, molecular reaction microscopes [9,10] (early nineties), retardation spectrometers [11,12,13] (early nineties), time projection chambers [14] (mid-seventies), and in muon [15], neutron [16,17], or nuclear decay [18] spectrometers, to name just a few experiments and surveys. In these applications charged particles are emitted over a wide range of emission angles ($0 < \theta \leq \pi/2$), and the number of orbits of gyration may vary widely ($0 < n < \infty$). This magnetic guidance of charged particles is the topic of the present paper.

For particles emitted from a *point* source, as sketched in Fig. 1, the distribution function of particle intensity over the detector plane is called a point spread function (PSF). Once a PSF is known, the particle distribution for any type of extended source can be calculated from it. Pictures like Fig. 1 are found in most introductory textbooks on physics, and one should think that the magnetic PSF for this setup is well known and no longer subject of investigation. But this is not the case, and we find some striking features in this PSF which, to our knowledge, have not been published before, and which may be of interest to a wider community.

Our particular interest is on the role of the magnetic PSF in the field of low-energy particle physics, which field is entering what some call the high-precision era [19,20,21], often searching for $10^{-4}$ effects that might signal new physics beyond the standard model, at a level where Monte-Carlo simulations often meet difficulties. In a recent publication [22] we had sketched the derivation of the PSF and its singularities in the context of neutron $\beta$-decay, and listed over a dozen neutron decay experiments that use magnetic guiding of charged reaction products for high precision measurements.

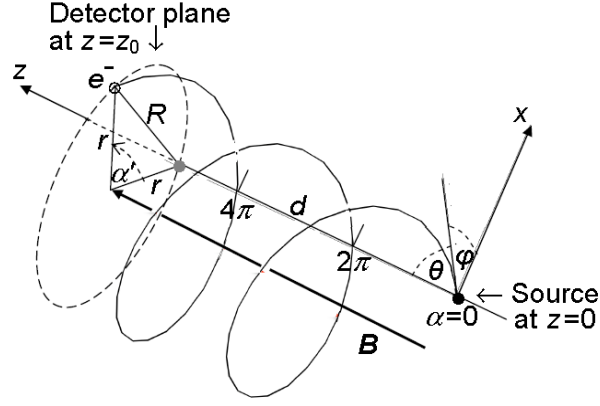

**Fig. 1.** An electron point source at $x = 0$ and a flat detector at $z = z_0$, both coupled by a uniform magnetic guide field $\boldsymbol{B}$ along $z$.

The paper is organized as follows: Section 2 sketches the conventional approach to the problem, which leads to a smooth hyperbolic PSF, used already some 30 years ago [7], and still in use up to these days [8]. Section 3 gives analytical proof that this conventional approach fails for any finite number $n$ of gyrations: At certain values $R_n$ of the particle's displacement $R$ on the detector (Fig. 1), strong resonances appear in the PSF, in spite of the wide angular range of emission angles $\theta$. Section 4 shows that this singular behaviour of the PSF makes the local response of a detector very sensitive to minute changes of instrumental parameters like field amplitude $B$, particle energy $E$, detector position $z_0$, or its angular adjustment. Section 5 extends these algebraic calculations to anisotropic sources and to non-uniform guide fields. Section 6 sketches a recent experiment [23] done at LANL, in which the resonances predicted in [22] were observed, and discuss some possible uses of the true magnetic PSF.

## 2. The conventional magnetic point spread function (PSF)

Without loss of generality, let the charged particles be monoenergetic electrons. Their radius of gyration $r$ depends on their polar angle of emission $\theta$ as

$$r = r_0 \sin\theta, \tag{1}$$

and on their kinetic energy $E$ via the maximum radius of gyration as

$$r_0 = p/eB = \sqrt{E(E+2mc^2)}/ecB, \tag{2}$$

for a field of amplitude $B$, with electron charge $e$, mass $m$, and relativistic momentum $p$. We assume a uniform magnetic field, for the non-uniform case see Sect. 5. The pitch of the helix

$$d = 2\pi r_0 \cos\theta \qquad (3)$$

is also indicated in Fig. 1. These formulas are found in most textbooks on electromagnetism.

Upon arrival of an electron on the detector, its total number of gyration orbits is $n' = z_0 / d$, where the slash reminds us that $n'$ needs not be an integer. The total phase angle of gyration is hence related to the angle of electron emission $\theta$ as

$$\alpha = 2\pi n' = z_0 / (r_0 \cos\theta). \qquad (4)$$

The smallest occurring phase angle, reached for electron emission under $\theta = 0$, is

$$\alpha_0 = z_0 / r_0 = 2\pi n_0, \text{ where} \qquad (5)$$

$$n_0 = eBz_0 / 2\pi p \qquad (6)$$

is the corresponding minimum number of orbits in the limit $\theta \to 0$ (where in fact a gyration is no longer visible). On the detector, the electron's point of impact (circular dot in Fig. 1) is displaced from its projected starting point (grey dot), reached for $\theta = 0$, by the distance

$$R = 2r |\sin \tfrac{1}{2}\alpha| = 2r_0 \sin\theta \, \sin \tfrac{1}{2}\alpha', \qquad (7)$$

where $\alpha' = (\alpha \text{ modulo } 2\pi)$ is the phase angle seen on the surface of the detector, with $0 \leq \alpha' \leq 2\pi$.

The conventional approach to the problem is to assume that all phase angles occur with the same probability, given by $dp_0/d\alpha = 1/\pi$, see [23] for a straightforward derivation of the magnetic PSF under this assumption. In this approach, the probability for finding an electron at displacement $R$ is

$$\frac{dp_0}{dR} = \left| \frac{dp_0}{d\alpha} \frac{d\alpha}{dR} \right| = \frac{1}{\pi} \left| \frac{d\alpha}{dR} \right|. \qquad (8)$$

Inserting, for a given $r$, the derivative $(dR/d\alpha)^{-1}$ as a function of $R$ from Eq. (7) leads to

$$\frac{dp_0}{dR} = \frac{2}{\pi\sqrt{4r^2 - R^2}}. \qquad (9)$$

For an isotropic source, integration of $dp_0/dR$ over angle of emission $\theta$ then gives a probability distribution that no longer depends on $R$,

$$g(R) \equiv \frac{dP}{dR} = \int_0^{\cos\theta_1} \frac{2}{\pi\sqrt{4r_0^2 \sin^2\theta - R^2}} d\cos\theta = \frac{1}{2r_0}. \qquad (10)$$

with the limit of integration $\cos\theta_1 = (1 - R^2/4r_0^2)^{1/2}$, where $P$ and $p_0$ are related as $p_0 = dP/d\cos\theta$.

For given $R$ and $dR$, the electrons arrive on the detector within an infinitesimal area of size $dA = 2\pi R\, dR$. The radially symmetric PSF $f(x,y) = f(R)$ on the detector surface then is the hyperbolic function

$$f(R) \equiv \frac{dP}{dA} = \frac{g(R)}{2\pi R} = \frac{1}{4\pi R r_0}, \tag{11}$$

with $R = (x^2+y^2)^{1/2}$. The singularity at $R = 0$ reflects the fact that all orbits cross the origin, for arbitrary values of emission angles $\theta$ and $\varphi$, and of energy $E$. Figure 2 of [24] shows a rough measurement of such a $1/R$ response. Note that in [22] we inadvertently called the function $g(R)$ the PSF, and not the function $f(R)$.

However, this conventional result cannot be the full truth, because, in its derivation from Eq. (7), the phase angle $\alpha$ and the pitch angle $\theta$ are treated as independent variables. In other words: for a given emission angle $\theta$, the electron on the detector is erroneously assumed to run on a circle through all values of $\alpha$.

## 3. Derivation of the true magnetic PSF

In reality both angles $\alpha$ and $\theta$ are uniquely linked to each other by Eqs. (4) and (5) as

$$\cos\theta = \alpha_0/\alpha. \tag{12}$$

After emission under $\theta$, the electron hence arrives at one fixed and predetermined position on the detector, given by the electron displacement

$$R(\alpha) = 2r_0\sqrt{1 - \alpha_0^2/\alpha^2}\,|\sin\tfrac{1}{2}\alpha|, \tag{13}$$

from Eq. (7). To increase the size of the phase angle $\alpha$ on the detector, one has to increase the emission angle $\theta$ at the source, and with it the gyration radius $r$ from Eq. (1), such that the trace on the detector is no longer a circle but some sort of a spiral.

Figure 2 shows $R$ in dependence of the total number of electron orbits $n' = \alpha/2\pi$. The function starts at the minimum number of orbits $n_0 = \alpha_0/2\pi$ for emission under $\theta = 0$, and continues to $\alpha/2\pi \to \infty$ for emission under $\theta = \pi/2$. One could as well write $R$ as a function of emission angle $\theta$, with the same final result for the PSF. However, while $\theta$ is the more directly accessible variable, the derivation is simpler in terms of the variable $\alpha$.

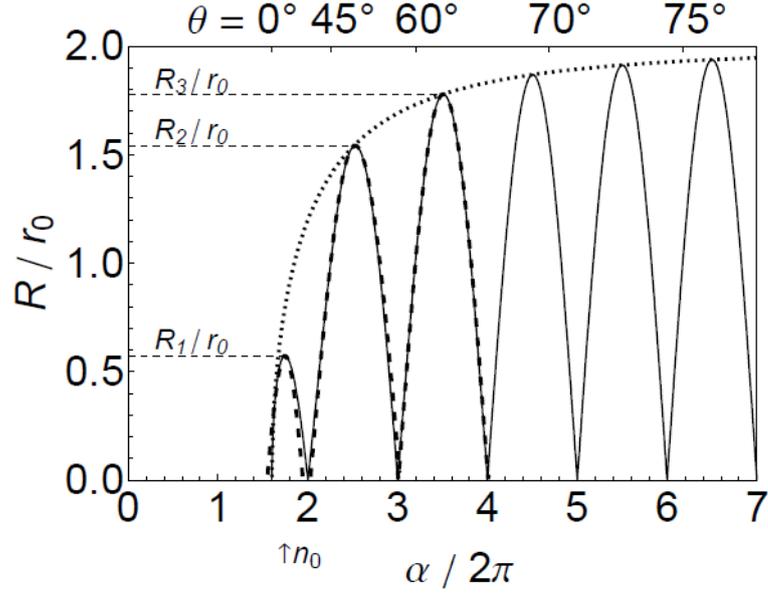

**Fig. 2.** Electron displacement $R$ on the detector from Eq. (13), plotted as a function of the number of orbits $n' = \alpha/2\pi$ (full line). The corresponding angles of emission $\theta$ from Eq. (4) are given on the upper axis. The envelope function $2r_0 \sin\theta(\alpha)$ in Eq. (7) is also shown (dotted line). The value $\alpha_0 = 10$ radians, or $n_0 = \alpha_0/2\pi = 1.6$ orbits, is the same as in the experiment [23], in which $B \sim \frac{1}{2}$ T, $z_0 \sim 0.1$ m, and $E \sim 1$ MeV, see Sect. 6. The dashed lines are the invertible approximations to $R(\alpha)$ from Eqs. (17) and (23).

We now come to the calculation of the true PSF. Often the polar angular distribution of the particles emitted from a source is developed in Legendre polynomials as functions of $\cos\theta$. Therefore the PSF is best written as

$$f(R) = \frac{1}{2\pi R}\left|\frac{dP}{d\cos\theta}\frac{d\cos\theta}{d\alpha}\frac{d\alpha}{dR}\right|. \tag{14}$$

We first treat the isotropic case $dP/d\cos\theta = 1$, for anisotropic sources see Sect. 5. Insertion of $d\cos\theta/d\alpha$ from Eq. (4) and of $(dR/d\alpha)^{-1}$ from Eq. (13) then leads to

$$f(\alpha) = \frac{1}{2\pi R r_0}\frac{\alpha_0(\alpha^2 - \alpha_0^2)^{1/2}}{\left|\alpha(\alpha^2 - \alpha_0^2)\cos\tfrac{1}{2}\alpha + 2\alpha_0^2 \sin\tfrac{1}{2}\alpha\right|}. \tag{15}$$

However, we need the PSF not as a function of $\alpha$, but as a function of $R$. Inversion of the multi-valued function $R(\alpha)$ of Fig. 2 to $\alpha(R)$ is the main obstacle to arriving at the true magnetic PSF.

While Eq. (15) is still exact, we now use the fact that the displacement $R(\alpha)$, Eq. (13), is the product of a rapidly varying function $|\sin(½\alpha)|$ and a slowly varying envelope $2r_0(1-\alpha_0^2/\alpha^2)^{1/2}$, as seen in the example of Fig. 2. Within a given cycle on the detector, numbered by the integer $n$, the slow envelope can therefore be piecewise approximated by a constant value $R_n$, which is best chosen to be the maximum of $R(\alpha)$ in the $n^{th}$ interval

$$R_n = \text{Max}[R(\alpha), 2\pi n \leq \alpha \leq 2\pi(n+1)], \qquad (16)$$

as indicated by the dashed horizontal lines in Fig. 2. The true function $R(\alpha)$ from Eq. (13) then is piecewise replaced by the approximate functions

$$R(\alpha) \approx R_n \cos[(\alpha - \alpha'_n)/2], \qquad (17)$$

each valid between $\alpha = 2\pi n$ and $\alpha = 2\pi(n+1)$, where $\alpha'_n$ is the position where $R(\alpha)$ obtains its maximum $R_n$. Note that for the lowest orbit starting at $n_0$, these equations hold only for integer $n_0$, for non-integer $n_0$ see below.

The dashed curves in Fig. 2 show these invertible functions Eq. (17), indicating the high quality of the approximation. For each orbit, the approximate $R(\alpha)$ can be resolved for $\alpha$,

$$\alpha_\pm(R) \approx \alpha'_n \mp 2\arccos(R/R_n). \qquad (18)$$

where $\alpha_+$ holds for the *rising* branches of $R(\alpha)$ in Fig. 2. This approximation holds for the *rising* branches of $R(\alpha)$ in Fig. 2, and $\alpha_-$ for the *falling* branches.

These approximate $\alpha$'s then must be inserted in Eq. (15). In this way one obtains for every cycle a partial PSF that we call $f_n$. These partial PSFs must then be summed up to obtain the magnetic PSF,

$$f(R) \approx \sum_{n=n_0}^{\infty} f_n(R). \qquad (19)$$

For integer $n_0$, one finds in the denominator of Eq. (15)

$$\cos(½\alpha) \approx \pm\sqrt{1-R^2/R_n^2}, \quad \sin(½\alpha) \approx R/R_n, \qquad (20)$$

with the plus sign for the rising branches, the minus sign for the falling branches. After Eqs. (20) are inserted in Eq. (15), one can, for not too small values of $n_0$, neglect $2\arcsin(R/R_n)$ in Eq. (18) for the other $\alpha$'s and set $\alpha(R) \approx 2\pi n$. The $n^{th}$ partial PSF for an isotropic source in a uniform magnetic field then is

$$f_n(R) \approx f_{n+}(R) + f_{n-}(R), \text{ with} \tag{21}$$

$$f_{n\pm}(R) = \frac{1}{2\pi R r_0} \frac{n_0(n^2 - n_0^2)^{1/2}}{\pm \pi n(n^2 - n_0^2)(1 - R^2/R_n^2)^{1/2} + n_0^2 R/R_n}. \tag{22}$$

Usually $n_0$ is not an integer, in which case special attention must be given to the lowest orbit, then numbered by the next-lower integer $n_f = \text{floor}(n_0)$. The width of the lowest interval in Fig. 2 then is less than unity, namely $2 - 1.6 = 0.4$. This requires replacing Eq. (16) by

$$R_{n_f} = \text{Max}[R(\alpha), 2\pi n_0 \leq \alpha \leq 2\pi(n_f + 1)], \tag{23}$$

and Eq. (17) by

$$R(\alpha) \approx R_{n_f} \left| \cos \frac{\alpha - \alpha'_{n_f}}{2(n_f + 1 - n_0)} \right|, \tag{24}$$

valid between $\alpha = 2\pi n_0$ and $\alpha = 2\pi(n_f + 1)$, where $\alpha'_{n_f}$ is the position where $R(\alpha)$ has its first maximum $R_{n_f}$. This can be resolved for $\alpha$ as

$$\alpha_\pm(R) \approx \alpha'_{n_f} \mp 2(n_f + 1 - n_0) \arccos(R/R_{n_f}). \tag{25}$$

Used in Eq. (15), this gives the first partial PSF $f_{n_f}(R)$, and the sum Eq. (19) starts at $n = n_f$. At the start of the lowest orbit where $\alpha$ is near $\alpha_0$, it is useful to replace Eq. (25) by the approximate inverted function $\alpha_+(R) \approx \alpha_0[1 + R^2/(8\sin^2 \alpha/2)]$. In Fig. 3b this is done up to $R = 0.34$. In Figs. 3a and 4 this replacement would avoid the little kinks seen at low $R$. - Finally, the normalization of $g(R)$ to unity was checked by numerical integration of $g(R)$, taken from the general Eq. (19).

## 4. Properties of the magnetic PSF

Figure 3a shows the new PSF $f(x,y)$, calculated directly from Eq. (15) with (18) and (25) for $n_0 = 1.6$. Strong resonances are seen whenever $R$ falls onto one of the maxima $R_n$ in Fig. 2. Summation of the $f_n$ is truncated at $n = 50$. With $\cos\theta = n_0/n$ from Eq. (12), this corresponds to a cut off angle $\theta = 88.2°$, with no visible effect to the PSF in Fig. 3a. Figure 2b shows $g(R) = 2\pi R f(R)$, both for the conventional and the new approach. The positions of the resonances along $R$ are well reproduced in our approximation. Even for $n_0 < 1$ well below one full orbit, the positions of the singularities are accurate, though their shapes are heavily distorted.

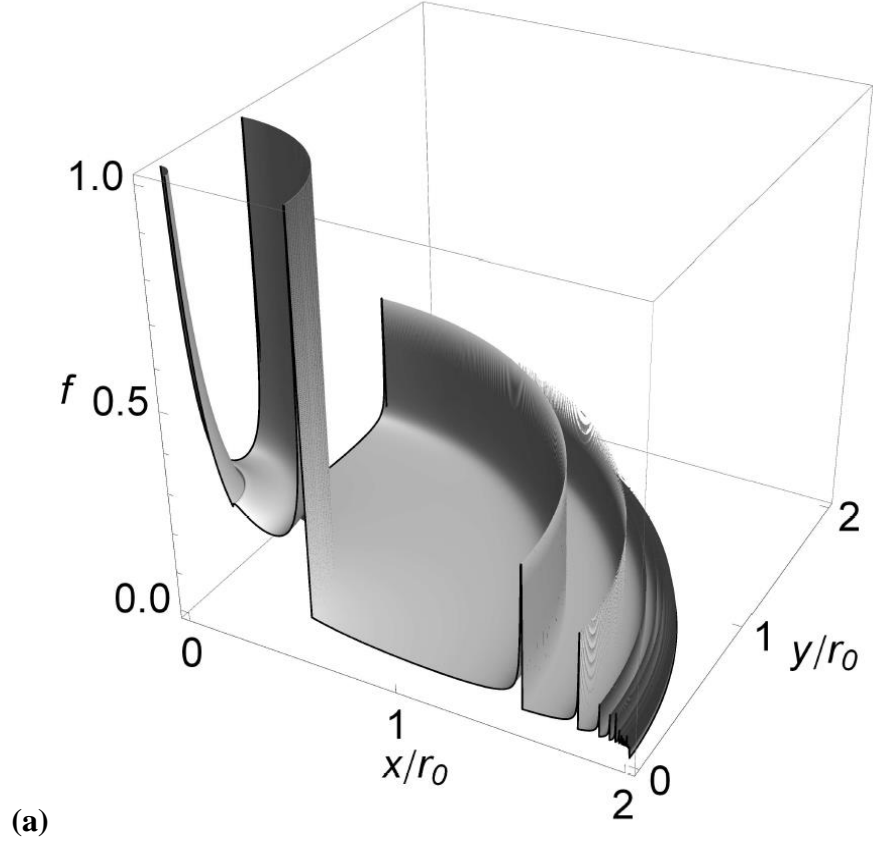

(a)

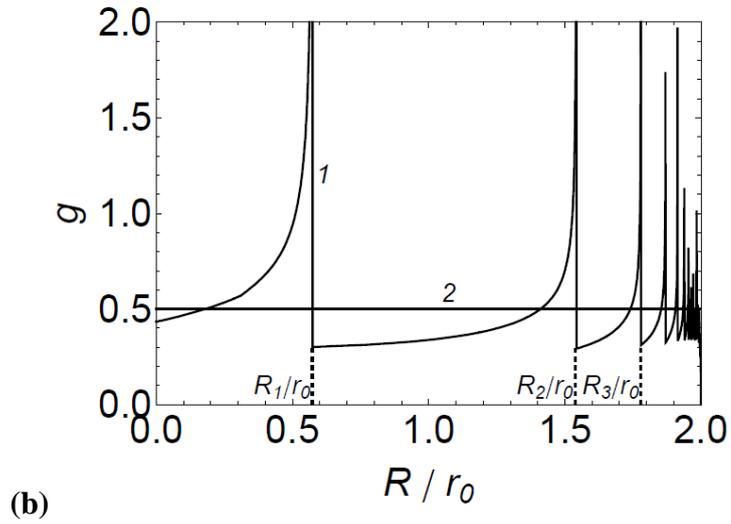

(b)

**Fig. 3. (a)** Magnetic point spread functions $f(x,y)$ from Eq. (15) with (18) and (25), in the first quadrant of the detector, for isotropic particle emission and uniform field, The same $\alpha_0 = 10$ radians, or $n_0 = 1.6$ orbits, are used as in Fig. 2. **(b)** The spiked function labelled 1 is the new probability distribution $g(R) = 2\pi R\, f(R)$. The constant function labelled 2 is the conventional distribution $g(R) = 1/(2r_0)$. The positions of the resonances (dashed vertical lines) coincide with the corresponding maxima $R_n$ in Fig. 2.

The positions and shapes of the resonances can be understood by looking at $R(\alpha)$ in Fig. 2. The fluctuating $R(\alpha)$ and its smooth envelope $|\sin\theta(\alpha)|$ coincide near each maximum $R_n$. This means that $\alpha$ and $\theta$ in Eq. (7) are strongly correlated there, contrary to the conventional assumption of independence of $\alpha$ and $\theta$. Therefore the deviations of the true PSF from the conventional PSF are strongest at $R \approx R_n$. They are singular because the derivative $d\alpha/dR$ in Eq. (14) diverges whenever $R(\alpha)$ reaches a maximum $R_n$ near a half-integer number of revolutions, where $R(\alpha)$ becomes stationary. If $R$ is increased above a particular $R_n$, the correlation between $\alpha$ and $\theta$ in the corresponding orbit is suddenly lost, cf. Fig. 2, and only the rather uncorrelated terms from the higher orbits contribute to Eq. (19). Therefore the true PSF falls steeply back to the conventional PSF whenever $R$ rises beyond one of the $R_n$.

In terms of emission angle $\theta$, the ring-shaped singularities in Fig. 3 occur at $\theta_1 = 27.6°$, $\theta_2 = 50.5°$, $\theta_3 = 63.0°$, etc., from Eq. (7). For large integer $n_0 \gg 1$, the first few singularities of the PSF occur for emission angles and displacements

$$\theta_n \approx \sqrt{2(n-n_0)/n_0}\,,\ R_n \approx 2r_0\sqrt{[2(n-n_0)+1]/n_0}\,. \tag{26}$$

Figure 4 demonstrates the high sensitivity of the true PSF to changes of external parameters that define the minimum number of orbits $n_0$, cf. Eq. (6). Note that in Fig. 4, $n_0$ is ten times larger than in Figs. 1 or 2. The sensitivity of the PSF to changes of $n_0$ can again be understood by looking at the corresponding changes of $R$ in Fig. 2. . For instance, when an integer $n_0$ changes continuously to the next higher integer $n_0+1$, then the first maximum moves from zero to its peak value (which is $R_1 = 0.73\,r_0$ for the interval from $n_0 = 1$ to $n_0 = 2$), and restarts at zero again. Note that for $n_0 \gg 1$, the first maximum of $R$ from Eq. (16) lies at $1/\sqrt{n_0}$, the second at $\sqrt{3/n_0}$, and $R_n \approx \sqrt{[2(n-n_0)+1]/n_0}$ for the first few maxima.

This parameter sensitivity of $n_0$ may average out the singularities of the PSF: These singularities are no longer individually resolved if $n_0$ changes by $\Delta n_0 > 1$. This happens when either $|\Delta B/B|$ (e.g., for extended sources), or $|\Delta z_0/z_0|$ (for axially extended sources), or $|\Delta p/p|$ (for continuous spectra) in Eq. (6) exceeds $1/n_0$. Note that often $\Delta n_0$ can be kept small by additional time-of-flight measurements and energy sensitive detection.

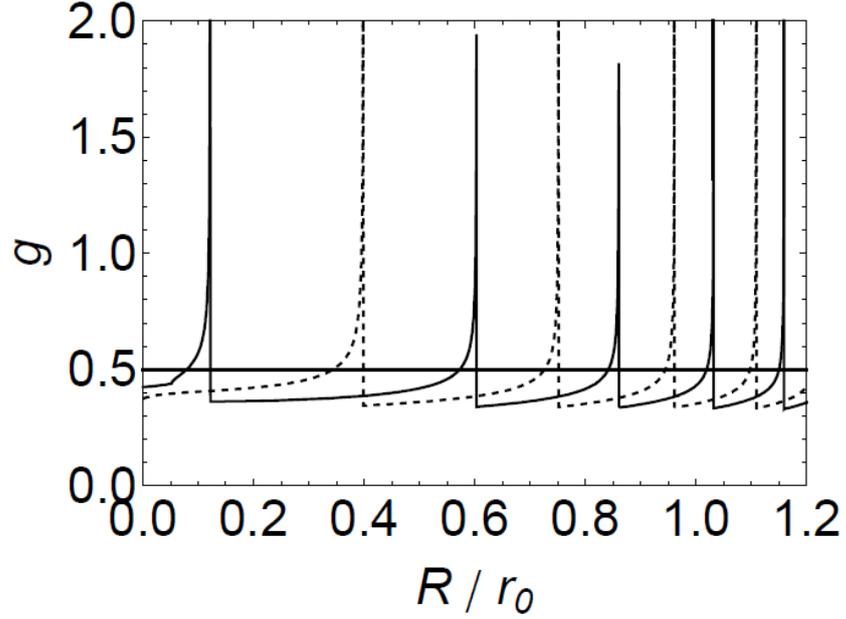

**Fig. 4.** Sensitivity of the probability distributions $g(R) = 2\pi R\, f(R)$ to changes of the external parameters, cf. Eq. (6). The conventional distribution (horizontal line) does not depend on external parameters. The true distribution (spiked curves) is highly parameter dependent. The full curve is for $\alpha = 105$ radians, or $n_0 = 16.7$ orbits, the dashed curve for a 3% lower value $\alpha = 102$ radians, or $n_0 = 16.2$ orbits.

## 5. Anisotropic sources and non-uniform guide fields

This section treats two extensions of our analytic PSF calculations. The first extension is to non-isotropic sources. Often the polar angular distribution $dP/d\cos\theta$ in Eq. (14) can be developed in associated Legendre polynomials, involving terms

$$\frac{dP}{d\cos\theta} \sim \cos^{l-m}\theta \, \sin^m\theta, \tag{27}$$

with integers $m \leq l$. An example is particle emission from atoms or nuclei that carry a vector or tensor polarization, see for instance chapters 19.3 and 20.5 in [25].

We begin with the conventional PSFs for anisotropic sources. If variations of $\Delta n_0$ are so large as to average out the singularities in the PSF, we can insert the distribution (27) into the integral Eq. (10). The conventional PSF for anisotropic sources then is

$$f_m^l(R) = g_m^l(R)/2\pi R, \text{ with} \tag{28}$$

$$g_m^l(R) = c_{l,m}\,(1/r_0)\,[1-R^2/4r_0^2]^{(l-m)/2}\, {}_2F_1(-\tfrac{1}{2}m, \tfrac{1}{2}(l-m+1); \tfrac{1}{2}(l-m+2); 1-R^2/4r_0^2), \tag{29}$$

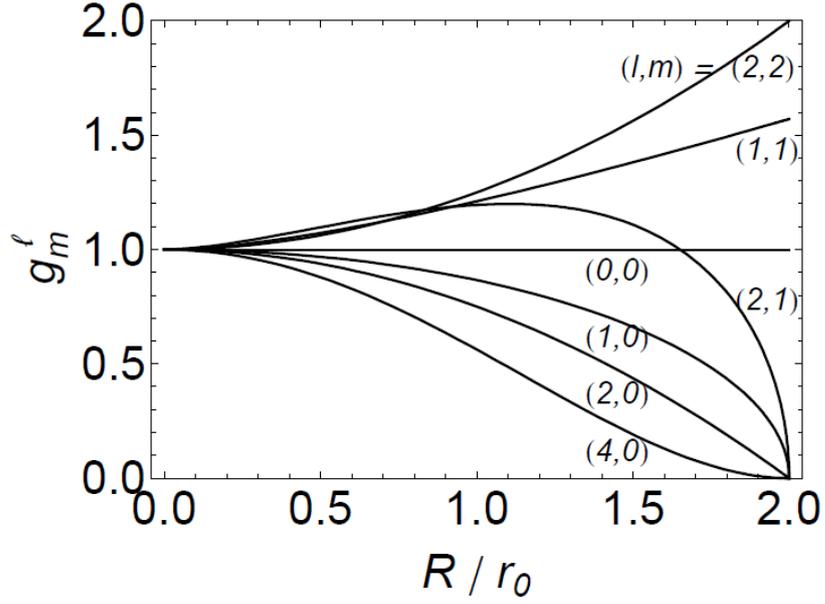

**Fig. 5.** Conventional probability distributions $g_m^l(R) = 2\pi R\, f_m^l(R)$ for anisotropic particle emission, from Eq. (29) for various values of $(l,m)$. The line $g_0^0(R) = \tfrac{1}{2}$ for an isotropic source is the same as the horizontal line in Fig. 4.

with the hypergeometric function ${}_2F_1$. The pre-factor

$$c_{l,m} = \frac{\sqrt{\pi}\, l\, \Gamma(\tfrac{1}{2}l)}{(l-m)\Gamma[\tfrac{1}{2}(l+m)]\, \Gamma[\tfrac{1}{2}(l-m)]}, \qquad (30)$$

with gamma functions $\Gamma$, is chosen such that $g_m^l(R) = 1$ at the origin $R = 0$.

Specific algebraic solutions for fixed values of $l$ and $m$ are best obtained directly from the integral Eq. (10) with $dP/d\cos\theta$ inserted, and not from the master equation (29) for arbitrary $l$ and $m$. For $m = 0$, that is, for ordinary Legendre polynomials involving only the terms $\cos^l\theta$, one finds

$$g_0^l(R) = (1 - R^2/4r_0^2)^{l/2}/2r_0. \qquad (31)$$

For $l = 0$ this coincides with the result (10) for isotropic sources. For $l = 1$ this last equation gives the PSFs for angular correlation functions, like the parity violating $\beta$ asymmetry, see [22] for details. Another example is particle emission from surfaces obeying Lambert's law. For $l = 1$, $m = 1$ we have $g_1^1(R) = E(1 - R^2/4r_0^2)$, with the complete elliptic integral $E(x)$, and so forth. Figure 5 shows $g_m^l(R)$ for various $l$ and $m$.

To obtain the true PSF for anisotropic sources, we insert $\cos\theta = n_0/n$ from Eq. (12) into Eq. (27) and multiply each partial PSF in the sum (19) by

$$\frac{dP}{d\cos\theta} \sim (n_0/n)^{l-m}[1-(n_0/n)^2]^{m/2}. \tag{32}$$

This gives the true PSFs $f_m^l(R)$, showing singularities for every $R = R_n$, rising over a base line that coincides with the corresponding conventional solution from Eq. (29).

Our second extension treats the case that the magnetic field, while still axially symmetric, is not uniform. In many experiments the field decreases continuously from $B$ at the source to $B'$ at the detector, which avoids glancing incidence on the detector for particles emitted near $\theta = \pi/2$. In the experiments [3-18] cited in the introduction, non-adiabatic transitions are strictly suppressed because their angle dependent energy losses would corrupt the measurements. In the adiabatic approximation, for $B' < B$ the inverse magnetic mirror effect makes the gyration radius widen from $r$ at the source to $r'$ on the detector, while the helix angle decreases from $\theta$ to $\theta'$, with

$$r' = r/b, \; \sin\theta' = b\sin\theta, \text{ with } b = \sqrt{B'/B}. \tag{33}$$

We expect that the usual adiabatic invariants, on which the above equations are based, guarantee that the particle distribution on the detector remains unchanged, stretched, however, by a factor $1/b$. While this conjecture sounds reasonable, we better check it analytically. To this end we set

$$u = \cos\theta, \; u' = \cos\theta', \tag{34}$$

both quantities being related to each other as

$$u = (1/b)\sqrt{b^2 - (1-u'^2)}. \tag{35}$$

The electron angular distribution then changes on the way from the source to the detector from $dP/du$ (taken, e.g., from Eq. (27)) to

$$\frac{dP}{du'} = \frac{dP}{du}\frac{du}{du'} = \frac{dP}{du}\frac{1}{b\sqrt{b^2 - (1-u'^2)}}. \tag{36}$$

Writing $\sin\theta' = (1-u'^2)^{1/2}$, for a decreasing field $b < 1$ this angular distribution narrows to $\theta' < \arcsin b = \theta'_c$, and diverges when the critical angle $\theta'_c$ is approached from below.

We next show that for a non-uniform guide field, relation (7) between the displacement $R'$ and the phase angle $\alpha'$ remains the same. First we note that the size of $R'$ is determined between the particles' last focus and the detector, separated from each other by the short distance δ$z$. This focus exists also for gyration in a non-uniform field, according to Busch's theorem [1]. The distance δ$z$ has at most the length of one pitch, $δz \leq d'$, which latter is of the order of the local gyration radius $d' \sim r'$, cf. Eq. (3). On the other hand, the adiabatic condition requires that variations of the field $B(z)$ are negligible over distances of one gyration radius $r'$, hence also over $δz \sim r'$. Therefore the field is uniform between the last focus and the detector, and Eq. (7), and with it Eq. (9), are valid also for $R'$ and $\alpha'$.

From this we conclude that the integral Eq. (10) can also be used for transport in non-uniform fields, but with the angular distribution Eq. (36) at the last focus inserted under the integral. The distribution function then reads

$$g'(R') = \int_{u'_1}^{u'_2} \frac{dP}{du} \frac{du}{du'} \frac{2u' du'}{\pi \sqrt{4 r_0'^2 (1-u'^2) - R'^2}} \tag{37}$$

with $r_0' = r_0 / b^2$ from Eq. (2), and with limits of integration $u'_1 = (1 - b^2)^{1/2}$ and $u'_2 = (1 - R^2 / 4 r_0'^2)^{1/2}$. When d$P$/d$u$ from Eq. (27) (written as a function of $u'$) is inserted, then indeed the master equation (29) is recovered, stretched by the factor $1/b = (B/B')^{1/2}$, and the PSF obeys

$$f'(R') = f(bR). \tag{38}$$

This is not surprising, because the integral (37) is mathematically equivalent (up to d$P$/d$u$) to the integral (10) for the uniform field case, both integrals differing merely by the substitution $u \rightarrow u'$, Eq. (35), and by the different length scales, $r_0' = r_0 / b^2$. This check of our conjecture was done for the conventional PSF, but holds for any distribution, including the true PSF.

## 6. Experimental verification, and possible applications of the new PSF

A recent experiment [23], done at the Los Alamos National Laboratory on the ultracold neutron decay spectrometer UCNA [26], has meanwhile confirmed the presence of resonances in the true PSF. In this work, an isotropic $^{207}$Bi conversion electron source ($E$ = 976 keV and 432 keV) was placed in a uniform magnetic guide field at 10 cm distance to a position sensitive Si detector of 10 cm diameter and 0.8 cm$^2$ pixel size. The field amplitude $B$ then was varied between 0.1 and 0.6 T, thus varying $r_0$ from 4.0 cm down to 0.7 cm. Every time one of the resonances of the PSF, shown in Fig. 3, entered or left a pixel, there was a sudden jump in the count rate from this pixel. Small changes of the magnetic field induced large changes in relative count rates (up to a factor of five), see figures 7 and 8 in [23], where the conventional PSF would predict a very smooth response. The agreement with simulated expectations is excellent. The Monte Carlo result in their Fig. 4 can be compared to our result in Fig. 2b, calculated with identical parameters, namely, $n_0 = 1.6$. Note that our resonances are narrower and steeper than the Monte Carlo result from [23]. Hence care must be taken when applying these calculations to possibly inherently broadened experimental data.

What are the possible benefits of having a new PSF? First, knowledge on the ring-shaped singularities in the PSF may promote understanding of experimental data and avoid surprises, for instance in reaction microscopy and similar experiments. Second, even when these rings remain unresolved, one must investigate their effect in high precision experiments, as was done for neutron decay in [22, 23]. Third, these rings may serve as an analytical tool to assess the proper working of magnetic guiding systems.

As an example for this last point we take the retardation spectrometer of the neutrino mass experiment KATRIN [13]. Its length from the effective $^3$H source to the detector is $z_0 = 51.8$ m. Although KATRIN's minimum field is well below 1 mT, its relevant average field is $\bar{B}_z = 1$ T, as deduced from [27]. For the 976 keV conversion line of a $^{207}$Bi test source, installed on-axis at the entrance of the spectrometer, Eq. (6) gives $n_0 = 2000$ orbits to the detector, all fully contained in the detector volume. Each conversion electron crosses the instrument axis at every cycle of gyration. KATRIN's ring-shaped electron detector, installed at the end of the instrument, can then measure the resulting PSF to check whether it is properly shaped. The Bi-source should be installed somewhat upstream of the initial field maximum to limit $\theta_{max}$, and the detector should be moved further downstream of the pitch field region to adapt gyration radii to detector size. In this way the entire spectrometer volume could be probed for possible discrepancies. These thoughts serve merely to remind the reader that new insights may generate new opportunities. (In the meantime I learned from C. Weinheimer that at 976 keV, adiabatic transport is no longer guaranteed in the KATRIN spectrometer.) Similar studies could be done on the neutron decay spectrometers PERC [28] with $n_0 = 200$, Nab [29] with $n_0 = 170$, or Perkeo-III [30] with $n_0 = 15$. Note added: A recent preprint [31] combines our initial approach with a special numerical method and finds results that coincide precisely with our result in Fig. 3b. The discrepancies mentioned in this note refer to the first version arXiv:1501.05131v2 [physics.ins-det] of the present preprint, where a less precise approximation had been used.

## 7. Conclusions

We calculated the point spread functions for charged particles in magnetic guide fields, which differ significantly from previously used results, as seen in Fig. 3. Algebraic results are derived for isotropic and anisotropic point sources, for uniform and non-uniform guide fields, valid also for sources rather close to the detector. The singularities found move rapidly across the PSF when the number of gyration orbits is changed by as little as a fraction of one orbit, see Fig. 4. A recent experiment done at LANL corroborates these results.

## Acknowledgements

This work was supported by the Priority Programme SPP 1491 of Deutsche Forschungsgemeinschaft. I thank L. Raffelt, B. Märkisch, F. Friedl, and H. Abele for helpful discussions on the applications of magnetic PSFs in neutron decay.

## References

[1] H. Busch, Berechnung der Bahn von Kathodenstrahlen im axialsymmetrischen elektromagnetischen Felde, Ann. Physik 386 (1926) 974.
[2] V. Kumar, Understanding the focusing of charged particle beams in a solenoid magnetic field, Am. J. Phys. 77 (2009) 737.
[3] C.M. Witcher, An electron lens type of beta-ray spectrometer, Phys. Rev. 60 (1941) 32.
[4] J.W.M. Dumond, An iron-free magnetic beta-ray spectrometer of high luminosity, resolving power, and precision for the study of decay schemes of the heavy isotopes, Ann. Phys. 2 (1957) 283.
[5] K. Siegbahn, β-ray spectrometer and design, in: K. Siegbahn, Ed., Alpha-, Beta- and Gamma-Ray Spectroscopy, p. 52, Elsevier, Amsterdam, 1955.
[6] G. Beamson, H.Q. Porter, D.W. Turner, Photoelectron spectromicroscopy, Nature 290 (1981) 556.
[7] P. Kruit, F.H. Read, Magnetic field paralleliser for $2\pi$ electron-spectrometer and electron-image magnifier, J. Phys. E 16 (1983) 313.
[8] R. Browning, Spatial resolution in vector potential photoelectron microscopy, Rev. Sci. Instr. 85 (2014) 033705.
[9] R. Moshammer, J. Ullrich, M. Unverzagt, W. Schmidt, P. Jardin, R.E. Olson, R. Mann, R. Dörner, V. Mergel, U. Buck, H. Schmidt-Böcking, Low-energy electrons and their dynamical correlation with recoil ions for single ionization of helium by fast, heavy-ion impact, Phys. Rev. Lett. 73 (1994) 3371.
[10] J. Ullrich, R. Moshammer, A. Dorn, R. Dörner, L.Ph.H. Schmidt, H. Schmidt-Böcking, Recoil-ion and electron momentum spectroscopy: reaction-microscopes, Rep. Prog. Phys. 66 (2003) 1463.

[11] G. Drexlin, V. Hannen, S. Mertens, C. Weinheimer, Current direct neutrino mass experiments, Adv. High Energy Phys. vol. 2013 (2013) article ID 293986.

[12] V.N. Aseev et al., Upper limit on the electron antineutrino mass from the Troitsk experiment, Phys. Rev. D 84 (2011) 112003.

[13] E.W. Otten, C. Weinheimer, Neutrino mass limit from tritium β decay, Rep. Prog. Phys. 71 (2008) 086201.

[14] J.N. Marx, D.R. Nygren, The time projection chamber, Phys. Today 31 (1978) 10.

[15] J. Kaulard, et al., Improved limit on the branching ratio of $\mu^+ \to e^-$ conversion on titanium, Phys. Lett. B 422 (1998) 334.

[16] H. Abele, The neutron. Its properties and basic interactions, Prog. Part. Nucl. Phys. 60 (2008) 1.

[17] D. Dubbers, M.G. Schmidt, The neutron and its role in cosmology and particle physics, Rev. Mod. Phys. 83 (2011) 1111.

[18] M. Beck, et al., First detection and energy measurement of recoil ions following beta decay in a Penning trap with the WITCH experiment, Eur. Phys. J. A 47 (2011) 45.

[19] K. Blaum, H. Müller, N. Severijns, Precision experiments and fundamental physics at low energies, Ann. Physik 525 (2013) A111.

[20] V. Cirigliano, S. Gardner, B. Holstein, Beta decays and non-standard interactions in the LHC era, Prog. Part. Nucl. Phys. 71 (2013) 93.

[21] A.S. Kronfeld, R.S. Tschirhart, Eds., Project X: physics opportunities, arXiv:1306.5009 [hep-ex].

[22] D. Dubbers, L. Raffelt, B. Märkisch, F. Friedl, H. Abele, The point spread function of electrons in a magnetic field, and the decay of the free neutron, Nucl. Instr. Meth. A 763 (2014) 112.

[23] S.K.L. Sjue, L. Broussard, M.F. Makela, P.L. McGaughey, Z. Wang, A.R. Young, B.A. Zeck, Radial distribution of charged particles in a magnetic field, Rev. Sci. Instr. 86 (2015) 023102.

[24] H. Kollmus, W. Schmitt, R. Moshammer, M. Unverzagt, J. Ullrich, A high resolution $4\pi$ multi-electron spectrometer for soft electrons, Nucl. Instr. Meth. B 124 (1997) 377.

[25] D. Dubbers, H.-J. Stöckmann, Quantum Physics: The Bottom-Up Approach - From the Simple Two-Level System to Irreducible Representations, Springer, Heidelberg 2013.

[26] M. P. Mendenhall, et al., Precision measurement of the neutron $\beta$-decay asymmetry, Phys. Rev. C 87 (2013) 032501.

[27] F. Glück, G. Drexlin, B. Leiber, S. Mertens, A. Osipowicz, J. Reich, Nancy Wandkowsky, Electromagnetic design of the large-volume air coil system of the KATRIN experiment, New J. Phys. 15 (2013) 083025.

[28] D. Dubbers, H. Abele, S. Baeßler, B. Märkisch, M. Schumann, T. Soldner, O. Zimmer, A clean, bright, and versatile source of neutron decay products, Nucl. Instr. Meth. A 596 (2008) 238.

[29] S. Baeßler, et al., Neutron beta decay studies with Nab, arXiv:1209.4663 [nucl-ex].

[30] B. Märkisch, H. Abele, D. Dubbers, F. Friedl, A. Kaplan, H. Mest, M. Schumann, T. Soldner, D. Wilkin, The new neutron decay spectrometer PERKEO III, Nucl. Instr. Meth. A 611 (2009) 216.

[31] H. Backe, Note: Precise radial distribution of charged particles in a magnetic guiding field, arXiv:1503.07064v1[physics.ins-det].